\documentclass[preprint,3p,sort&compress]{elsarticle}
\usepackage{amsmath}	
\usepackage{amssymb}

\usepackage[usenames,dvipsnames,svgnames,table]{xcolor}
\usepackage{graphicx}
\usepackage{dcolumn}
\usepackage{bm}
\usepackage{hyperref}
\makeatletter

\def\ps@pprintTitle{%
 \let\@oddhead\@empty
 \let\@evenhead\@empty
 \def\@oddfoot{\centerline{\thepage}}%
 \let\@evenfoot\@oddfoot}
\makeatother

\usepackage{tikz}
\usetikzlibrary{positioning,decorations.pathreplacing,shapes}
\usetikzlibrary{calc, shapes, backgrounds}
\usetikzlibrary{intersections}
\usetikzlibrary{arrows,decorations.pathmorphing,backgrounds,positioning,fit,petri}
\usetikzlibrary{decorations.markings}
\usetikzlibrary{calc}

\newcommand{\NGen}{\ensuremath {n_G}}
\newcommand{\TF}{\ensuremath {\textcolor{RoyalBlue}{T_F}}}
\newcommand{\CA}{\ensuremath {\textcolor{RoyalBlue}{C_A}}}
\newcommand{\CF}{\ensuremath {\textcolor{RoyalBlue}{C_F}}}

\def\ag{\ensuremath{a_s}}
\def\at{\ensuremath{a_t}}
\def\al{\ensuremath{a_{\lambda}}}
\def\tr{\ensuremath{\mathrm{tr}}}

\def\MS{\ensuremath{\overline{\mathrm{MS}}}}

\def\bgV{\ensuremath{\hat V}}

\journal{Physics Letters B}

\begin{document}
\begin{frontmatter}


\title{Four-loop strong coupling beta-function in the Standard Model}

\author[a]{A.V.~Bednyakov}
\address[a]{Joint Institute for Nuclear Research,\\
  141980 Dubna, Russia}
\ead{bednya@theor.jinr.ru}
\author[b]{A.F.~Pikelner\fnref{fn1}}%
\fntext[fn1]{On leave of absence from Joint Institute for Nuclear
  Research, 141980 Dubna, Russia}
\ead{pikelner@theor.jinr.ru}
\address[b]{%
 II. Institut f\"ur Theoretische Physik,                            
    Universit\"at Hamburg, \\
  Luruper Chaussee 149, 22761 Hamburg, Germany
}%


\begin{abstract}
  In this letter we present our results for the four-loop beta-function of 
  the strong coupling in the Standard Model of fundamental interactions. 
  The expression is obtained from gluon self-energy diagrams in the background field gauge
  without application of special infra-red rearrangement tricks.
  We take top-Yukawa and self-Higgs interactions into account, but neglect electroweak gauge couplings.
  Ambiguities due to $\gamma_5$ treatment are discussed and a particular  
  ``reading'' prescription for odd Dirac traces is advocated.
\end{abstract}


\end{frontmatter}

Most of viable models describing Nature at high energies are based on gauge symmetries.
Quantum chromodynamics (QCD) is a gauge theory of strong interactions and it is important to study its strength described by the coupling $\alpha_s$ 
both in the low- and high-energy limit. 
At low energies the QCD interactions play a dominant role in binding quarks and gluons together into nucleons. 
At larger scales the coupling $\alpha_s$ decreases \cite{Gross:1973id,Politzer:1973fx}
due to non-Abelian nature of the underlying gauge theory.
Nevertheless, the precise value of $\alpha_s$ is of paramount importance both for modern and future colliders and for theoretical studies of physics going beyond the Standard Model (SM) of fundamental interactions.

The renormalization group equations (RGE) relate couplings at different scales. By solving them one can not only confront measurements carried out at different energies with theory but also study asymptotic behavior of the latter at scales inaccessible in current and even future experiments.

The progress in calculation of beta-functions --- key RGE quantities --- is tightly connected to the  
introduction of dimensional regularization \cite{'tHooft:1973mm} and (modified) minimal  subtraction (or $\MS$) scheme. 
The former does not break gauge symmetry in $d=4 - 2 \epsilon$ dimensions and 
the advantage of the latter lies in the fact that beta-functions are extracted only from ultraviolet (UV) asymptotics of Feynman integrals. This fact allows one to drastically simplify a calculation by modifying the infra-red (IR) structure of the considered integrals by means of the so-called infra-red rearrangement (IRR) procedure~\cite{Vladimirov:1979zm}.

Pure QCD results for the strong coupling beta-function are known for quite a long time up to four loops~\cite{Gross:1973id,Politzer:1973fx,
Caswell:1974gg,Jones:1974mm,Egorian:1978zx,
Tarasov:1980au,Larin:1993tp,
vanRitbergen:1997va,Czakon:2004bu}. The four-loop results are
obtained with the help of IRR procedure leading to four-loop vacuum
diagrams with all lines having the same mass. The beta-function is extracted
from the quark-gluon \cite{vanRitbergen:1997va} and
ghost-gluon \cite{Czakon:2004bu} vertex renormalization constants and the corresponding
wave function renormalization constants. Independently, the ghost field 
renormalization constant and the ghost-gluon vertex renormalization
constant were calculated in Ref.~\cite{Chetyrkin:2004mf}. The quark field
renormalization constant \cite{Chetyrkin:1999pq} was found by a different method bringing 
the problem to the calculation of three-loop massless propagator-type diagrams.

It is obvious that in a precise study of QCD processes at high energies one should consider other SM interactions and their effect on the running of $\alpha_s$.
Recently, the full set of three-loop beta-functions for all SM parameters, including the strong coupling, was found in a series of papers \cite{Mihaila:2012fm,Bednyakov:2012rb,Bednyakov:2012en,Chetyrkin:2013wya,Bednyakov:2013eba,Bednyakov:2014pia,Bednyakov:2013cpa}.

In this letter we present our result for the dominant four-loop contribution to the
beta-function of the strong coupling in the SM.  
In our calculation we neglect
the electroweak gauge interactions, but keep top-Yukawa and Higgs
self-interactions along with the well-known QCD corrections. The
calculation is carried out in the background-field gauge (BFG)~\cite{Abbott:1980hw,Denner:1994xt}.
The advantage of BFG lies in the QED-like relation between the 
gauge coupling renormalization constant $Z_{a_{s}}$ and that of the background gluon field $Z_{\bgV_3}$:

\begin{equation}
  \label{eq:bgf}
  Z_{a_{s}} = 1/Z_{\bgV_3},\qquad Z = 1 + \sum\limits_{i=1}^{\infty} \frac{\delta Z^{(i)}}{\epsilon^i}.
\end{equation}

This allows us to
obtain the final result by considering massless propagator-type
integrals. 
For the present calculation we evaluate the required four-loop integrals contributing to the two-point Green function for the background gluon field
and exploit multiplicative renormalizability of theory as
in Refs.~\cite{Larin:1993tp,Pickering:2001aq,Velizhanin:2008rw}. 
We only need to calculate bare Green-functions up to the $l$-loop level and re-express all the model parameters in
terms of the renormalized ones in the $\MS$-scheme:
\begin{equation}
  \Gamma_{\mathrm{ren}}^{(l)}=Z_{\Gamma}^{(l)}\left[1+ \sum\limits_{k=1}^l\Gamma_{\mathrm{B}}^{(k)}
    (a_{\mathrm{B}}) \right],\;
  a_{\mathrm{B}}=Z_aa_{\mathrm{ren}}\label{eq:mulren},
\end{equation}
	where $a$ collectively denotes the SM couplings in the considered limit, i.e., strong - $g_s$, top-Yukawa - $y_t$, and that of Higgs self-interaction - $\lambda$, together with $SU(3)$ 
	gauge-fixing parameter $\xi$:  
\begin{equation}
	(16 \pi^2) a = \left\{ g_s^2, y_t^2, \lambda, (16 \pi^2) \xi \right\}.
	\label{eq:coupldef}
\end{equation}

As it is clear from \eqref{eq:mulren},  due to finiteness of
$\Gamma_{\mathrm{ren}}$, RHS should be finite too and at each order of perturbation theory we have an equation on $Z_{\Gamma}$ terms. 
Given the knowledge of all three-loop renormalization constants, it is possible to extract the four-loop contribution 
to the renormalization constant $Z_{\bgV_3}$ for the background gluon field $\bgV_3$.

For diagram generation 
the package
\texttt{DIANA}~\cite{Tentyukov:1999is}, which internally
uses~\texttt{QGRAF}~\cite{Nogueira:1991ex}, was utilized. 
After some color \cite{vanRitbergen:1998pn} and Dirac algebra all the generated two-point functions were mapped onto three auxiliary topologies, each containing 11
propagators and 3 irreducible numerators.
Before the actual integration-by-parts (IBP) \cite{Chetyrkin:1981qh} reduction, a set of reduction rules was prepared by means of
\texttt{LiteRed}\cite{Lee:2012cn} package. After that the prepared set 
was passed to the \texttt{C++} version of
\texttt{FIRE} package~\cite{Smirnov:2014hma}, which allows one to reduce the obtained four-loop integrals in parallel on a multithread machine.
The IBP reduction leads to a small set of master integrals. The expressions for the latter are 
known in analytical form up to the finite parts \cite{Baikov:2010hf}.
The master integrals were also cross-checked numerically in Ref.~\cite{Smirnov:2010hd} and some additional terms 
in $\epsilon$-expansion were found in Ref.~\cite{Lee:2011jt}.
We also perform some independent evaluation of several three-loop integrals. 
The simplest four-loop integrals with three-loop
insertions in a one-loop integral were checked by means of the \texttt{FORM}-based package
\texttt{MINCER}~\cite{Gorishnii:1989gt,Larin:1991fz}.

It is also worth mentioning that as an independent cross-check of our setup, we prepared a simple
QCD model with additional fermion in the adjoint representation of SU(3) color group (``gluino'').
The beta-function for such a model at the three-loop order can be predicted 
by means of proper color factor substitutions~\cite{Clavelli:1996pz}. 
At four loops similar procedure ceases to be sufficient and some additional information from the direct calculations \cite{Chetyrkin:1996ez} of
QED-type diagrams is required to predict the beta-function.
We compared the predicted results \cite{Pikelner:2015} with that calculated by means of the above-mentioned
setup and found a perfect agreement.

After these kind of tests we address the issue of finding the SM four-loop contribution to the gauge coupling beta-function. 
The calculation of bare Green-functions requires evaluation of traces
over Dirac matrices in $d\neq 4$ dimensions and an additional complication, comparing to QCD,
arises when traces involving $\gamma_5$ matrices are present (see, e.g., Ref.~\cite{Jegerlehner:2000dz}).

One needs to be careful in maintaining $\gamma_5$ anticommutativity and strict four-dimensional relation
\begin{equation}
  \tr\left( \gamma^\mu \gamma^\nu \gamma^\rho \gamma^\sigma \gamma_5 \right) = -4 i \epsilon^{\mu\nu\rho\sigma}, 
  \label{eq:gamma5_trace}
\end{equation}
involving totally antisymmetric tensor with $\epsilon^{0123}=1$. 
At three loops \cite{Chetyrkin:2012rz,Bednyakov:2012en} it was proven by direct calculations that it is possible to use semi-naive 
approach and utilize both $\{\gamma^\mu,\gamma_5\}=0$ and \eqref{eq:gamma5_trace} without paying much attention to apparent non-cyclicity of the trace operation in $d\neq 4$.
The formal $\epsilon$-tensor originating from the trace \eqref{eq:gamma5_trace} can only give a non-trivial contribution 
if two such traces are present and the anti-symmetric tensors are contracted by means of 
\begin{equation}
  \epsilon^{\mu\nu\rho\sigma} \epsilon_{\alpha\beta\gamma\delta}
  = - {\mathcal T}{}^{[\mu\nu\rho\sigma]}_{[\alpha\beta\gamma\delta]},
  \qquad
  {\mathcal T}{}^{\mu\nu\rho\sigma}_{\alpha\beta\gamma\delta}=
  \delta^\mu_\alpha
  \delta^\nu_\beta
  \delta^\rho_\gamma
  \delta^\sigma_\delta.
  \label{eq:eps_contaction}
\end{equation}
	Strictly speaking, the Kronecker delta-symbols in \eqref{eq:eps_contaction} should be considered as four-dimensional objects and the contraction
    with the remaining part of a diagram should be carried out after subtraction of infinities via $R$-operation \cite{Bogoliubov:1957gp}.
    However, it is not convenient in a massive calculation involving thousands of Feynman diagrams and it is tempting to use $d$-dimensional $\delta^\mu_\nu$ satisfying $\delta^{\mu}_{\mu} = d$ in the bare $d=4-2 \epsilon$ theory. 
    It is easy to convince oneself that the cyclic property of traces gives rise to an ambiguity $\mathcal{O}(\epsilon)$, which can play a role in determining RG coefficients. 
    However, it turns out that at the three-loop level a non-trivial contribution originating from the contraction \eqref{eq:eps_contaction} appears for the first time in the Yukawa coupling beta-function~\cite{Chetyrkin:2012rz}. 
    Both gauge-coupling and higgs self-coupling turn out to be free from this kind of contributions due to gauge-anomaly cancellation conditions fulfilled with the SM.
	
    In our calculation we tried to employ the above-mentioned semi-naive approach to study the diagrams giving rise to a non-zero terms due to \eqref{eq:eps_contaction}. 
     A typical diagram is shown in Fig.~\ref{fig:gam5eps}. 
    By counting coupling constants and performing color algebra it is easy to convince oneself that
     $\gamma_5$ affects only $\ag^2 \at^2 T_F^2$ contribution with $T_F=1/2$.
     We have 24 planar diagrams of this type and 48 non-planar graphs, which can be 
     obtained by permutation of internal lines connecting two fermion traces. 
It contains box-type closed sub-loops involving two gauge bosons
     and two scalars as external legs~\footnote{ 
Similar sub-diagrams of three-loop Yukawa vertex were discussed in papers~\cite{Chetyrkin:2012rz,Bednyakov:2012en} in the same context. }.
Direct evaluation of the diagram shows that, indeed, the resulting expression  is free from higher poles in $\epsilon$. 

An additional argument to the fact that the four-loop higher poles are not affected by different $\gamma_5$ prescriptions comes from an observation that they can be  found in advance from the known three-loop results \cite{Mihaila:2012fm,Bednyakov:2012rb,Bednyakov:2012en,Chetyrkin:2013wya,Bednyakov:2013eba,Bednyakov:2014pia,Bednyakov:2013cpa} and the so-called pole equations~\cite{'tHooft:1973mm}.
Dangerous contributions due to  \eqref{eq:eps_contaction} to the four-loop higher poles can only appear if three-loop Yukawa coupling beta-function $\beta_{a_t}$ is involved.
However, it is easy to prove that it is not the case for $\beta_{a_s}$. 
Since one-loop gauge-coupling beta-functions does not depend on other couplings, only one- and two-loop terms in $\beta_{a_t}$  contribute to the four-loop pole equations for $a_s$ and we expect no dangerous high-order poles in $Z_{a_s}$.
It is worth stressing that the argument can also be applied to all gauge-coupling beta-functions in the full SM provided that all gauge anomalies are canceled.

This kind of reasoning lead us to a premature conclusion that the 
semi-naive approach is sufficient to get the correct answer for the strong coupling beta-function at four loops. The question whether the absence of higher poles in the considered diagrams is sufficient for them to be unambiguous was initially left without consideration.  

However, a more careful study, triggered by the appearance of similar result in Ref. \cite{Zoller:2015tha}, 
of all the above-mentioned 72 diagrams have shown that our initial treatment leads to results inconsistent with formal charge-conjugation symmetry. 
For example, the left fermion loop in Fig.\ref{fig:gam5eps} gives
\begin{equation}
	\tr \left(\hat p_1 \gamma_\rho \hat p_2 \hat p_3 \gamma_5 \hat p_4 \gamma_\mu \right)
	=
        \tr \left(\hat p_1 \gamma_\rho \hat p_2 \hat p_3 \gamma_5 \hat p_4 \gamma_\mu \right)^T	
	,
    \label{eq:fermline}
\end{equation}
	where $\mu,\rho$ are Lorentz indices of gluons connected to the loop, $p_i$ correspond to momenta of fermion propagators, and $\gamma_5$ comes from the vertex with neutral would-be goldstone boson ($\chi$). 
	Assuming standard relations $C\gamma_\mu C^{-1} = - \gamma_\mu^T$ and  $C\gamma_5 C^{-1} = \gamma_5^T$ due to charge conjugation transformation $C$, one can show that the right-hand side (RHS) of Eq.~\eqref{eq:fermline} can be rewritten in the form 
\begin{equation}
    \tr \left(\gamma_\mu (-\hat p_4) \gamma_5 (-\hat p_3)   (-\hat p_2) \gamma_\rho (-\hat p_1) \right),
    \label{eq:flipfermline}
\end{equation}
	 which corresponds to a diagram with flipped fermion flow of the considered closed chain. 
	 As a consequence, these two diagrams should be equal. 
	 However, an inconsistent treatment of $\gamma_5$ could spoil this $C$-symmetry. 
    As it is know\cite{Korner:1991sx}, the cyclic property of the Dirac trace should be abandoned if one tries to utilize both $\gamma_5$ anticommutativity and trace relation \eqref{eq:gamma5_trace}.
    Due to this, the ``reading" point \cite{Korner:1991sx}, which defines the position of $\gamma_5$ in an ``odd" trace, should be chosen consistently for diagrams related by $C$-symmetry. 
    In our initial treatment the choice was made by the utilized diagram generation code \texttt{DIANA}, which
    starts writing closed fermion loops from a propagator with certain momentum, irrespectively of its fermion flow. e.g., from $(-\hat p_1)$ for Eq.~\eqref{eq:flipfermline}. 
    As it will be shown latter, this prescription leads to an inconsistent result.
    
In order to shed light on possible solutions to this issue, i.e., we study different ``reading" points \cite{Korner:1991sx} for the involved Dirac traces.
We ``cut" a trace at a certain point, i.e., start writing a fermion string from either propagator or vertex, 
and $\{\gamma_5,\gamma_\mu\}=0$, $\gamma_5^2=1$ properties to anticommute $\gamma_5$ to the \emph{rightmost} position of the corresponding chain.
For example, Eq.~\eqref{eq:fermline} corresponds to cutting the trace at the external gauge vertex, while the flipped version \eqref{eq:flipfermline} 
is cut at the propagator entering the vertex.
The resulting traces involve more than four $\gamma$-matrices and direct application of the trace condition \eqref{eq:gamma5_trace} is not possible. We made use of the algorithm given in the FORM \cite{Kuipers:2012rf} manual to 
reduce these traces to Eq.~\eqref{eq:gamma5_trace}.   

Simple ``scan" over different reading points indicates that, indeed, there is an ambiguity in the resulting expression depending on the position, at which the traces are ``cut". 
We can distinguish three situations: A) both traces start (or end\footnote{We can also consider situations when traces are terminated at fermion propagators, but it is easy to convince oneself that these cases are also included in our consideration.}) at external vertices; B) only one trace starts(or ends) at an external vertex; C) both traces start(or end) at some internal vertex.
The ambiguous divergent contribution to the gluon self-energy from a diagram involving odd traces with $\gamma_5$ can be  parameterized in the following way
\begin{equation}
\frac{\ag^2 \at^2 T_F^2}{\epsilon} 
\left( X_1 + X_2 \zeta_3 \right).
\label{eq:ambiguity}
\end{equation}
	For the case A, the planar graphs give $X_1=1/6$, $X_2=0$, while non-planar ones lead to $X_1=-1/18$ and $X_2=1/6$. Summing contributions from 24 planar and 48 non-planar diagrams we obtain
\begin{equation}
\frac{\ag^2 \at^2 T_F^2}{\epsilon} 
\left( \frac{4}{3} + 8 \zeta_3 \right).
\label{eq:g5_ext_vertex}
\end{equation}

For the case B we have found that the corresponding coefficients are multiplied by a factor of 2, while the prescription C give rise to a factor of 3. 

It is worth mentioning that we also tried to utilize Larin-like \cite{Larin:1993tq} prescription for dealing with $\gamma_5$, i.e., the substitution
\begin{equation}
	 \gamma_\mu \gamma_5 \to -\frac{i}{3!}\epsilon_{\mu\nu\rho\sigma} \gamma^\nu \gamma^\rho \gamma^\sigma,
    \label{eq:larin}
\end{equation}
	which, in a cyclic trace with anticommuting $\gamma_5$,  can be interpreted as a reading prescription, which corresponds to an average of two adjacent cut-points   
\begin{equation}
\tr\left(\ldots\gamma_\mu \gamma_5\right) \to 
\frac{1}{2} \left[ 
	  \tr\left(\ldots \gamma_\mu \gamma_5\right)
    - \tr\left(\gamma_\mu \ldots \gamma_5\right) 
    \right].
\end{equation}
This latter fact was confirmed by direct calculations.
There is a subtlety with \eqref{eq:larin} coming from the fact that $\gamma_\mu$ may not only come from a gauge vertex, but from a propagator, connecting external and internal vertices. For example, if one terminates the left fermion chain in Fig.~\ref{fig:gam5eps} by the external vertex, but the right one --- by the Yukawa vertex involving, e.g., a neutral Goldsone boson $\chi$ (marked by a blue circle), the outcome will be the mean of the results corresponding to the cases A and B, i.e., $X_1 = 1/2 \cdot 1/6 \cdot  (1_A + 2_B) = 1/4$.  

Similar situation appears if we \emph{do not} move $\gamma_5$ in (a part of) the trace with single $\gamma_5$, but just use \eqref{eq:larin} without anticommuting it to some cut-point. This approach turns out to be equivalent to the one, when the cut-point is actually fixed to be the vertex, in which $\gamma_5$ appears. Again, referring to Fig.~\ref{fig:gam5eps} with internal $\chi$, application of \eqref{eq:larin} from the very beginning gives rise to $X_1 = 1/2 \cdot 1/6 \cdot ( 2_{B}  + 3_{C}) = 5/12$. 
The only complication in the charged $\phi^\pm$ goldstones case comes from the necessity of additional averaging over four different ways to choose the position of single $\gamma_5$ in both traces, since $1\pm\gamma_5$ enters each vertex with $\phi^\pm$. For example,
\begin{equation}
X_1 = \frac{1}{2^2} \cdot \frac{1}{2^2} \cdot \frac{1}{6} \cdot \left( 2_B \cdot 4
+ 3_C \cdot 3\cdot 4 
        \right) = \frac{11}{24}, 
\end{equation}
However, it is easy to understand that such a prescription will give rise to a result incompatible with the above-mentioned C-symmetry, since flipping the right loop fermion flow results in
\begin{equation}
X_1 = \frac{1}{2^2} \cdot \frac{1}{2^2} \cdot \frac{1}{6} \cdot \left( 3_C \cdot 4 \cdot 4 \right) = \frac{1}{2}. 
\end{equation}
Nevertheless, it is worth pointing that the issue arises only due to identical handling of both diagrams.  
In general, one is able to use different reading prescriptions for different diagrams and by a proper choice 
it is possible to maintain the C-symmetry. We can parametrize our ignorance of the correct reading prescription  
by means of a function
\begin{equation}
R(x,y,z) = \frac{1_A \cdot x + 2_B \cdot y + 3_C \cdot z}{x + y + z},\qquad 1\leq R(x,y,z) \leq 3, \qquad x,y,z \in \mathbb{N},
	\label{eq:Rambiguity}
\end{equation}
	which corresponds to an average over $x+y+z$ reading points, 
	$x$ of which are of type A, $y$ - of type B, and $z$ - of type C.


\tikzset{
  di/.style={line width=1pt,draw=red, postaction={decorate},
    decoration={markings,mark=at position .55 with
      {\arrow[draw=blue]{>}}}},
  pl/.style={line width=1pt,draw=blue, postaction={decorate},
    decoration={markings,mark=at position .55 with
      {\arrow[draw=black]{>}}}},
  photon/.style={line width=1pt,decorate, decoration={snake}, draw=green},
  gluon/.style={decorate, draw=magenta,
    decoration={coil,amplitude=3pt, segment length=3.5pt}},
  higgs/.style={line width=1pt,draw=blue,dashed}
}  
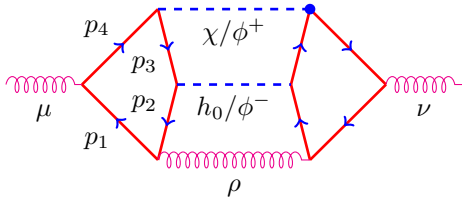
\begin{figure}[th]
  \centering
  \begin{tikzpicture}
    \coordinate (c1) at (-0.75,0); \coordinate (c2) at (0.75,0);
    \draw[di] (-1,1) -- (c1) node[midway,below left] {$p_3$}; 
    \draw[di] (c1) -- (-1,-1) node[midway,above left] {$p_2$}; 
    \draw[di] (-1,-1) -- (-2,0) node[midway,below left] {$p_1$}; 
    \draw[di] (-2,0) -- (-1,1) node[midway,above left] {$p_4$};
    \draw[di] (1,-1)--(c2); 
    \draw[di] (2,0) -- (1,-1); 
    \draw[di] (1,1) -- (2,0);
    \draw[di] (c2) -- (1,1) node[fill=blue,circle,inner sep=0pt,minimum size=4pt] {}; 
    \draw[gluon] (-3,0) -- (-2,0) node[midway,below=4pt] {$\mu$}; 
    \draw[gluon] (2,0) -- (3,0) node[midway, below=4pt] {$\nu$};
    \draw (0,1) node[anchor=north] {$\chi/\phi^+$}; 
    \draw[higgs] (-1,1) -- (1,1); 
    \draw ($(c1)!0.5!(c2)$) node[anchor=north] {$h_0/\phi^-$}; \draw[higgs] (c1) -- (c2); 
    \draw[gluon] (-1,-1)  -- (1,-1) node[midway,below=4pt] {$\rho$};
  \end{tikzpicture}
  \caption{A typical diagram with two fermion traces giving rise to a non-zero contribution to $\beta_{a_s}$
  at four loops due to the appearance of $\gamma_5$ in Yukawa-type vertices of the would-be Goldstone bosons $\chi,\phi^\pm$. Fermion line momenta $p_i$ are indicated. The momentum flow in the loop coincides with the fermion flow. The latter is denoted by arrows.}
  \label{fig:gam5eps}
\end{figure}

In spite of the observed ambiguity, in what follows we provide some arguments leading us to a conclusion that the prescription $C$ with $R=3$ is preferred among the others.

First of all, as it is noted in Ref.~\cite{Korner:1991sx}, reading points at external vertices 
could spoil gauge invariance of the final result of our two-point function. Indeed, we have checked that
finite parts of the (sum of) corresponding diagrams are not transverse if both Dirac traces are cut at the external vertices. It turns out that the prescription B also yields zero upon multiplication by the product of external momenta $q_\mu q_\nu$. Nevertheless, if we want to avoid non-symmetric treatment of the external vertices, we are left with the only reasonable prescription, which is C in our case.   

One more argument comes from utilization of Fierz identities, which convert two closed fermion chains into a single loop. 
A well-known subtlety lies in 
the fact that the relations in non-integer dimensions involve infinitely many terms (see, e.g., Ref.~\cite{Blatter:1992yr}). 
In our calculation we have tried to incorporate the following identity 
\begin{eqnarray}
\sum\limits^{\infty}_{\mu=1} 
\gamma^\mu_{a_1a_4}  
\gamma^\mu_{a_3a_2} & = &
2^{-d/2} \sum\limits_{l=0}^{\infty} (-1)^\frac{l(l+1)}{2} \frac{(d-2l)}{l!}\sum\limits_{\mu_i=1}^{\infty}\Gamma^{\mu_1,...,\mu_l}_{a_1a_2}
   \Gamma^{\mu_1,...,\mu_l}_{a_3a_4},
   \label{eq:fierz}
\end{eqnarray}
in which totally anti-symmetric combinations
\begin{equation}
\Gamma^{\mu_1,...,\mu_l} = \frac{1}{l!}\sum\limits_P (-1)^P
\gamma_{\mu_{i_1}}  
\gamma_{\mu_{i_2}}  
...
\gamma_{\mu_{i_l}}  
\end{equation}
are introduced. In our case the two contracted indices in RHS of \eqref{eq:fierz} correspond to the gluon propagator connecting two fermion traces in the Feynman gauge. 
We have checked that up to terms $l=4$ the obtained result coincide with the one corresponding to $R=3$.    

Finally, we re-calculated the problematic diagrams by applying an infra-red rearrangement technique 
\cite{Vladimirov:1979zm,Misiak:1994zw}, i.e., by transforming the integrals into fully-massive bubbles. 
An important difference from Ref.~\cite{Zoller:2015tha} lies in the fact that we perform Dirac traces 
applying self-consistent BMHV-algebra
\cite{'tHooft:1973mm,Breitenlohner:1977hr}  
\emph{after} (tensor) integrals are evaluated in $d$-dimension.
We have used \texttt{spinney}  package \cite{Cullen:2010jv} to keep track of the dimension of the involved Lorentz indices and
cross-checked that the final expression does not depend on the cut point.

After providing this kind of arguments, let us now proceed with our result. We define the four-loop beta-function for $a_s$ as
\begin{equation}
  \label{eq:betadef}
  \frac{d\;\ag}{d\;\log{\mu^2}}=\beta_{\ag} \ag= - \ag \sum\limits_{i=0}^3\beta_ih^{i+2}
\end{equation}
where we use $h$ to count the powers of coupling constants given in \eqref{eq:coupldef}. 
Our final result for $\beta_3$ 
can be written in terms of SU(3) casimirs and the number of SM generations - $\NGen$: 
  \begin{eqnarray}
    \label{eq:beta4sm}
    \beta_3 & = &  \beta_3^{\rm QCD}(n_f = 2 \NGen) 
    %
    %
                  + \ag^3\at \left[ 
                  \TF\CF^2\left(   6 - 144 \zeta_3\right ) + 
                  \TF\CA\CF\left( \frac{ 523}{9} - 72\zeta_3\right) +
                  \frac{1970}{9}\TF\CA^2 \right. \nonumber \\
            & - &
                  \left.\frac{1288}{9}\TF^2\CF \NGen -  
                  \frac{872}{9}\TF^2\CA \NGen \right]
                %
                %
             + \ag \at^3 \TF\left (
                  \frac{423}{2}
                 + 12 \zeta_3
                 \right)
                 + 60 \ag \at^2\al \TF
                 - 72 \ag \at\al^2 \TF\nonumber\\
                  & - &  \ag^2\at^2 \left[
                  \TF^2\left( 48 - 96 \zeta_3 +   
		  \textcolor{RedOrange}{\underbrace{R}_{3}}\cdot \left[
                  \textcolor{RedOrange}{\frac{16}{3}+ 32\zeta_3}
                  \right]
                  \right)+
                  \TF\CF\left( 117 - 144\zeta_3\right)+
                  222 \TF\CA \right],
                %
                %
                %
  \end{eqnarray}
where pure $n_f$-flavour QCD contribution $\beta _3^{\rm QCD}$ is available
from Refs.~\cite{vanRitbergen:1997va,Czakon:2004bu}, in which the same notation for
beta-function and coupling constants was used.
The result is free from gauge-fixing parameter dependence. This serves as a welcome check since at the intermediate steps one has
to take into account that the gauge-fixing parameter $\xi$ is renormalized in the same way as quantum gluon field.
In \eqref{eq:beta4sm} we emphasize terms $\mathcal{O}(\ag^2 \at^2 \TF^2)$, which are affected by contributions from diagrams similar to that given in Fig.~\ref{fig:gam5eps} and multiply the corresponding terms by the factor $R$
from Eq.~\eqref{eq:Rambiguity}.

\begin{figure}[th]
	     \centering
	    \includegraphics[width=0.48\textwidth]{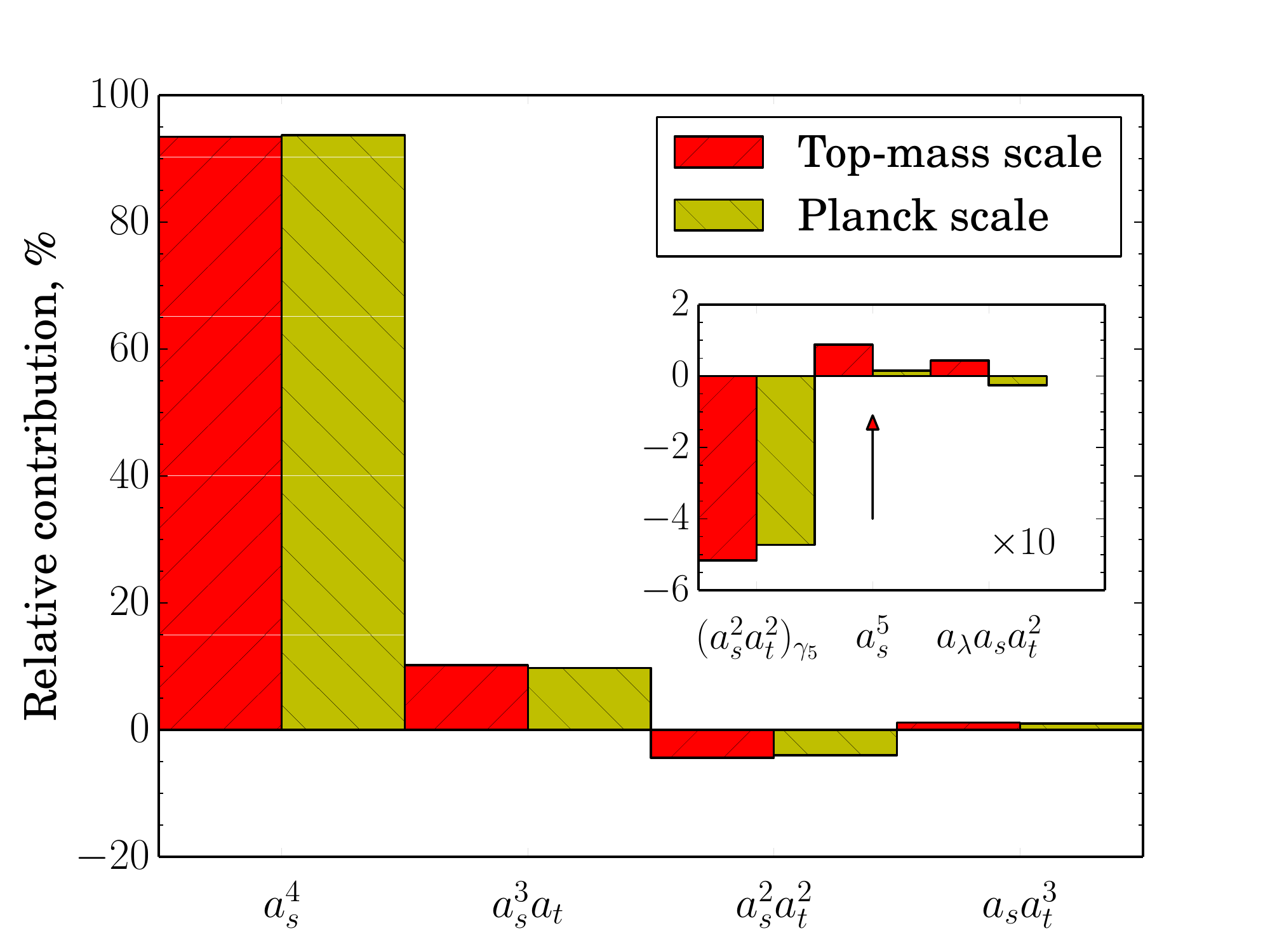}
	    \caption{Relative size of the calculated four-loop
              contributions and the pure QCD five-loop term
              $\mathcal{O}(\ag^5)$ (found recently in
              Ref.~\cite{Baikov:2016tgj} and indicated by arrow)
		     with respect to their sum. Nontrivial part due to $\gamma_5$ is also indicated. The numbers in the inset should be multiplied by 10.
	     	Negligible correction due to $\ag\at\al^2$ is not shown.}
	        \label{fig:rel_contrib}
\end{figure}

Now we turn to the numerical impact of the addition of new terms (see Fig.~\ref{fig:rel_contrib}). Using as input the values of strong, Yukawa and Higgs self-coupling
from\cite{Kniehl:2015nwa} at the scale $\mu=M_t$ and also evolving the
same input upto the Planck scale, we see that at both scales the leading pure QCD contribution $\mathcal{O}(\ag^4)$ accounts $\sim 93\%$ of the total four-loop result while the dominant $\mathcal{O}(\ag^3 \at)$ terms due to top-Yukawa only account
$\sim 10\%$. We observe that the $\mathcal{O}(\ag^2 \at^2)$
contribution has an opposite sign and is about $\sim 5\%$ of $\beta_3$ but terms $\mathcal{O}(\ag \at^3)$ have the same sign and contribute to the order of $\sim 1\%$. The other
contributions are less than $1\%$ at both the considered scales. Terms due to non-trivial $\gamma_5$ treatment lead to about $-0.5\%$ reduction in $\beta_3$. 
Nevertheless, the corresponding contribution is much larger than pure
QCD five-loop corrections, found recently in
Ref.~\cite{Baikov:2016tgj} and partially confirmed in Ref.~\cite{Luthe:2016ima}.

In conclusion, we calculated four-loop beta-function for the SM strong coupling in the limit of vanishing electroweak gauge-couplings.
By means of the background-field gauge formalism the result was obtained solely by considering four-loop massless propagator-type diagrams. 
We studied the $\gamma_5$-issue and found that the semi-naive treatment is \emph{not} sufficient to obtain unambiguous expression.
In our approach the ambiguity manifests itself in the dependence of the result on the ``reading" prescription, chosen to evaluate traces with odd number of $\gamma_5$, and was parametrized by the factor $1\leq R \leq 3$ in our final expression for $\beta_3$~\eqref{eq:beta4sm}.
Nevertheless, we provide several arguments supporting the prescription with $R=3$, for which both traces are ``cut" at some internal vertex.

The obtained result is a necessary step towards a four-loop RG analysis of the SM.
However, one should keep in mind that proper inclusion of three-loop non-QCD decoupling corrections 
is required to evaluate $\alpha_s$ consistently at the electroweak scale in the full SM (for pure QCD results up 
to the four-loop level see Refs.~\cite{Chetyrkin:1997sg,Schroder:2005hy,Kniehl:2006bg} and for 
the leading two-loop electroweak corrections see Ref.~\cite{Bednyakov:2014fua}).
In addition, the results can be applied to the study of the so-called Weyl consistency conditions \cite{Osborn:1989td,Jack:1990eb,Osborn:1991gm} 
within the SM, thus, extending the analysis of Ref.~\cite{Antipin:2013sga} to the four-loop level.
\subsection*{Acknowledgments}

We thank V.N. Velizhanin, O.L. Veretin and A.I. Onischenko for fruitful discussions. The correspondence with R.N. Lee and A.V. Smirnov regarding \texttt{LiteRed} 
and \texttt{FIRE} is kindly appreciated. 
We also would like to thank M. Zoller, who performed completely independent calculation of the same quantity in Ref.\cite{Zoller:2015tha} and drew our our attention to the remaining ambiguity, unnoticed in the first version of this paper. 
This work was supported in part by DFG Grant No.\ SFB~676, MES of Russia Grant No.\ MK--1001.2014.2, the Heisenberg-Landau Programme, and the Dynasty
Foundation. 
\bibliographystyle{elsarticle-num}
\bibliography{as4lsm}

\end{document}